\documentclass[a4paper]{PoS}

\title{Three little radio galaxies in the early Universe}

\ShortTitle{Three little radio galaxies in the early Universe}

\author{\speaker{K. \'E. Gab\'anyi},$^{ab}$ S. Frey,$^b$ Z. Paragi,$^c$ H. Cao,$^{d,e}$ T. An,$^f$ L. I. Gurvits,$^{c,g}$ T. Sbarrato,$^h$ K.~Perger,$^{i,b}$ K. Rozgonyi$^j$ and Gy. Mez\H{o}$^b$\\
\llap{$^a$}MTA-ELTE Extragalactic Astrophysics Research Group, Budapest, Hungary\\
\llap{$^b$}MTA CSFK, Konkoly Observatory, Budapest, Hungary\\ 
\llap{$^c$}Joint Institute for VLBI ERIC, Dwingeloo, the Netherlands\\
\llap{$^d$}Xinjiang Astronomical Observatory, Chinese Academy of Sciences, Urumqi, China \\
\llap{$^e$}School of Electronic and Electrical Engineering, Shangqiu Normal University, Shangqiu, Henan, China\\
\llap{$^f$}Shanghai Astronomical Observatory, Chinese Academy of Sciences, Shanghai, China\\
\llap{$^g$}Department of Astrodynamics and Space Missions, Delft University of Technology, Delft, the Netherlands\\
\llap{$^h$}Universit\`a degli Studi di Milano-Bicocca, Milano, Italy\\
\llap{$^i$}Department of Astronomy, E\"otv\"os University, Budapest, Hungary\\
\llap{$^j$}International Centre for Radio Astronomy Research, Perth, Australia\\
E-mail: \email{krisztina.g@gmail.com}, \email{frey.sandor@csfk.mta.hu}}

\abstract{Volonteri et al. (2011) found that the number of radio-loud quasars above redshift 4 calculated from the luminosity function (based upon {\it Swift}/BAT observations) is much smaller than the number estimated from the known high-redshift beamed sources, blazars, assuming that for every beamed source with a Lorentz factor of $\Gamma$, statistically $2 \Gamma^2$ non-beamed sources should exist. To explain the missing misaligned (non-beamed) population of high-redshift sources, they proposed various explanations, involving heavy optical obscuration and significantly different Lorentz factors at early cosmological epochs. Our EVN observations targeting high-redshift ($z>4$) blazar candidates revealed 3 sources not showing relativistic beaming, but rather kpc-scale double structures. These three sources have significant radio emission resolved out with the EVN, while they are compact on $\sim 5-10$ arcsec scale. 
Our dual-frequency ($1.5$ and $5$\,GHz) e-MERLIN observations of these three sources revealed a rich morphology, bending jets, and hot spots with possible sites of interaction between the jets and the surrounding medium at intermediate scales.}

\FullConference{14th European VLBI Network Symposium \& Users Meeting (EVN 2018)\\
		8-11 October 2018\\
		Granada, Spain}

\newcommand{\farcs}{\mbox{\ensuremath{.\!\!^{\prime\prime}}}}

\begin{document}

\section{Introduction}
Blazars are radio-loud active galactic nuclei whose jets point very close to the line of sight. The jets are Doppler-boosted leading to e.g., apparent superluminal motion and brightness temperature exceeding the equipartition \cite{equipartition} or the inverse Compton limit \cite{inverseCompton}. For every blazar (beamed sources) whose jets point toward us at a viewing angle $\lesssim 1/\Gamma$ (where $\Gamma$ is the jet plasma bulk Lorentz factor), statistically there must exist $\sim~2\Gamma^2$ number of sources whose jets have larger viewing angles, thus non-blazar (or misaligned) radio-loud sources \cite{Volonteri2011}. The luminosity function derived from the {\it Swift}/BAT observations of blazars was used in \cite{Volonteri2011} to calculate the expected number density of high-mass blazars between redshifts $2$ and $6$ by assuming different $\Gamma$ values. They found that from $z>3$ onward, the number of radio-loud quasars derived by cross-matching the Sloan Digital Sky Survey (SDSS, \cite{sdss_dr7}) and the Faint Images of the Radio Sky at Twenty-Centimeters survey (FIRST, \cite{FIRST_White}) is much smaller than the expected number estimated from the high-redshift blazars detected by {\it Swift}/BAT at high energies when assuming $\Gamma=15$.

We conducted European VLBI Network (EVN) observations of several high-redshift ($z>4$) sources to study their parsec-scale properties. In \cite{Cao2017}, out of the four targets which were claimed to be blazars based upon their X-ray characteristics and radio loudness, two (J2220$+$0025 and J1420$+$1205) exhibited arcsec-scale double structures. In \cite{Coppejans2016}, out of the ten targeted $z>4.5$ sources, five showed definite blazar-like characteristics, while one source (J1548$+$3335) found to be a double with a separation of $\sim 800$\,mas. In all three sources, from the two components observed at $1.7$\,GHz only one can be detected at $5$\,GHz. This result indicates that a compact radio-emitting core and a steep-spectrum hot spot along the jet in the lobe was revealed in the EVN observations. 

In the following, we assume a $\Lambda$CDM cosmological model with the $H_0=70$\,km\,s$^{-1}$\,Mpc$^{-1}$, $\Omega_\mathrm{m}=0.27$, and $\Omega_\Lambda=0.73$. The spectral index, $\alpha$ is defined as $S\sim\nu^\alpha$, where $S$ is the flux density, $\nu$ is the frequency.

\section{e-MERLIN observations and data reduction}

To investigate the arcsec-scale characteristics of the three sources, we conducted e-Multiple Element Remotely Linked Interferometer Network (e-MERLIN) observations at 1.5 GHz and at 5 GHz (project code: CY5211). The 1.5-GHz observations took place on 2017.05.15 and 2017.06.30. The 5-GHz observations took place on 2017.05.13, 2017.06.26 and 2017.06.27. In all observing segments, the array consisted of the Jodrell Bank Mk2, Pickmere, Darnhall, Knockin, Cambridge and Defford antennas. At $1.5$\,GHz and $5$\,GHz, eight and four spectral windows were used, respectively, each with $128$ channels. The total bandwidth was $512$\,MHz in both bands. Observations were conducted in phase-reference mode. The on-source integration times and phase reference calibrator sources are listed in Table \ref{tab:obs}.

Data reduction was done with the Common Astronomy Software Applications (CASA, \cite{casa}), using the e-MERLIN pipeline\footnote{www.e-merlin.ac.uk/data\_red/tools/eMCP.pdf} with extensive help from the e-MERLIN science team (J. Moldon).

\begin{table}
\centering
\begin{tabular}{ccccc}
\hline
Target name & Redshift & \multicolumn{2}{c}{On-source time (h)} & Phase-reference source \\
 & & $1.5$\,GHz & $5$\,GHz  \\
\hline
J1420$+$1205 & $4.034$ & $7.2$ & $14.9$ & J1415$+$1320\\
J1548$+$3335 & $4.680$ & $16.2$ & $17.9$ & J1544$+$3240 \\
J2220$+$0025 & $4.205$ & $9.6$ & $14.3$ & J2219$+$0229 \\
\hline
\end{tabular}
\caption{Details of the e-MERLIN observations.}
\label{tab:obs}
\end{table}

\section{Results}

For all three sources, we detected with e-MERLIN at $1.5$\,GHz and also at $5$\,GHz the components revealed by our earlier $1.7$-GHz EVN observations \cite{Cao2017,Coppejans2016}. 

\subsection{J1420$+$1205}

\begin{figure}
\begin{minipage}[t]{.54\textwidth}
\centering
\vspace{0pt}
\includegraphics[bb=25 280 490 760, clip=, width=\textwidth]{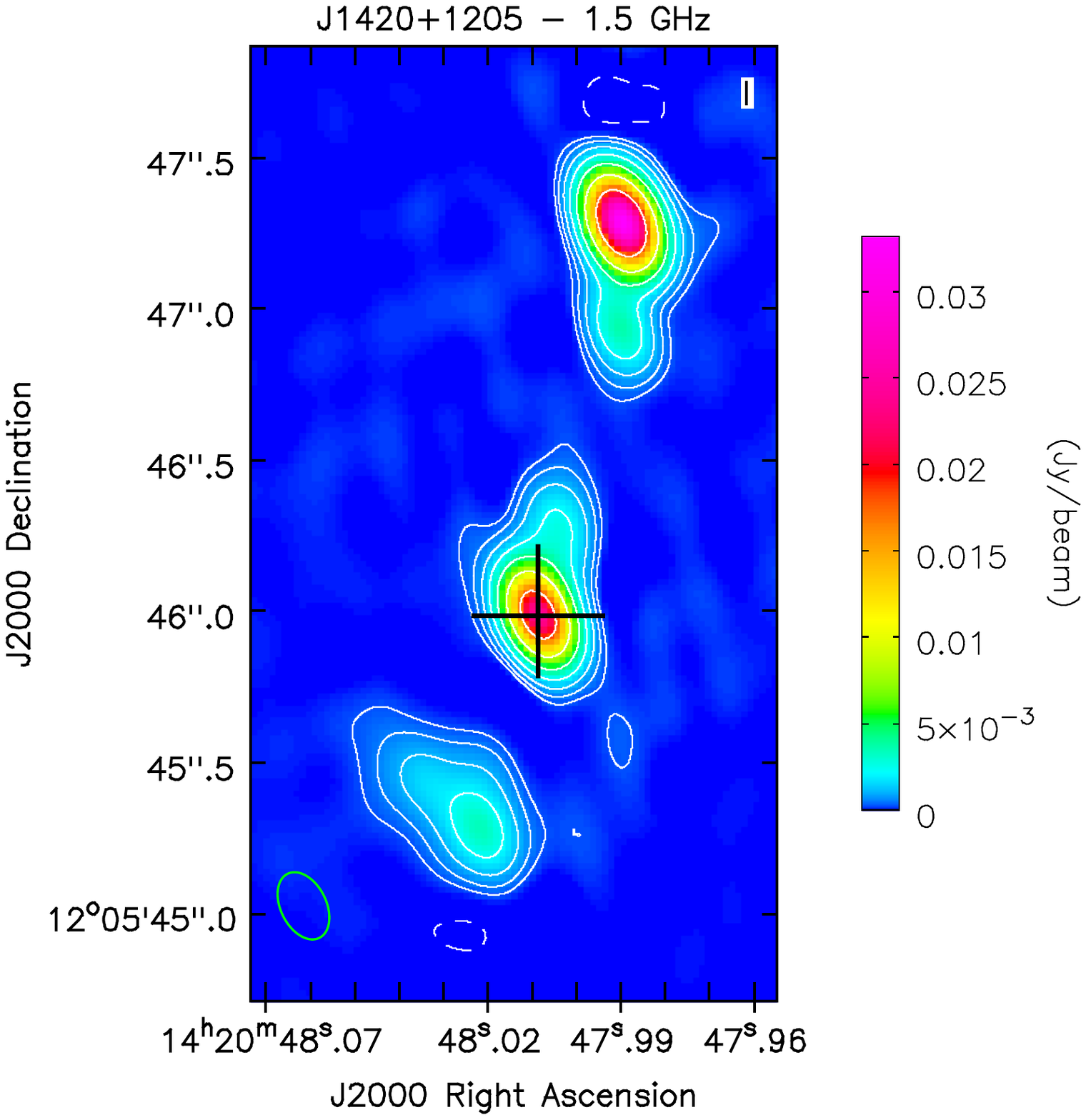}
\end{minipage}
\hfill
\begin{minipage}[t]{.46\textwidth}
\centering
\vspace{0pt}
\includegraphics[bb=25 240 490 800, clip=, width=\textwidth]{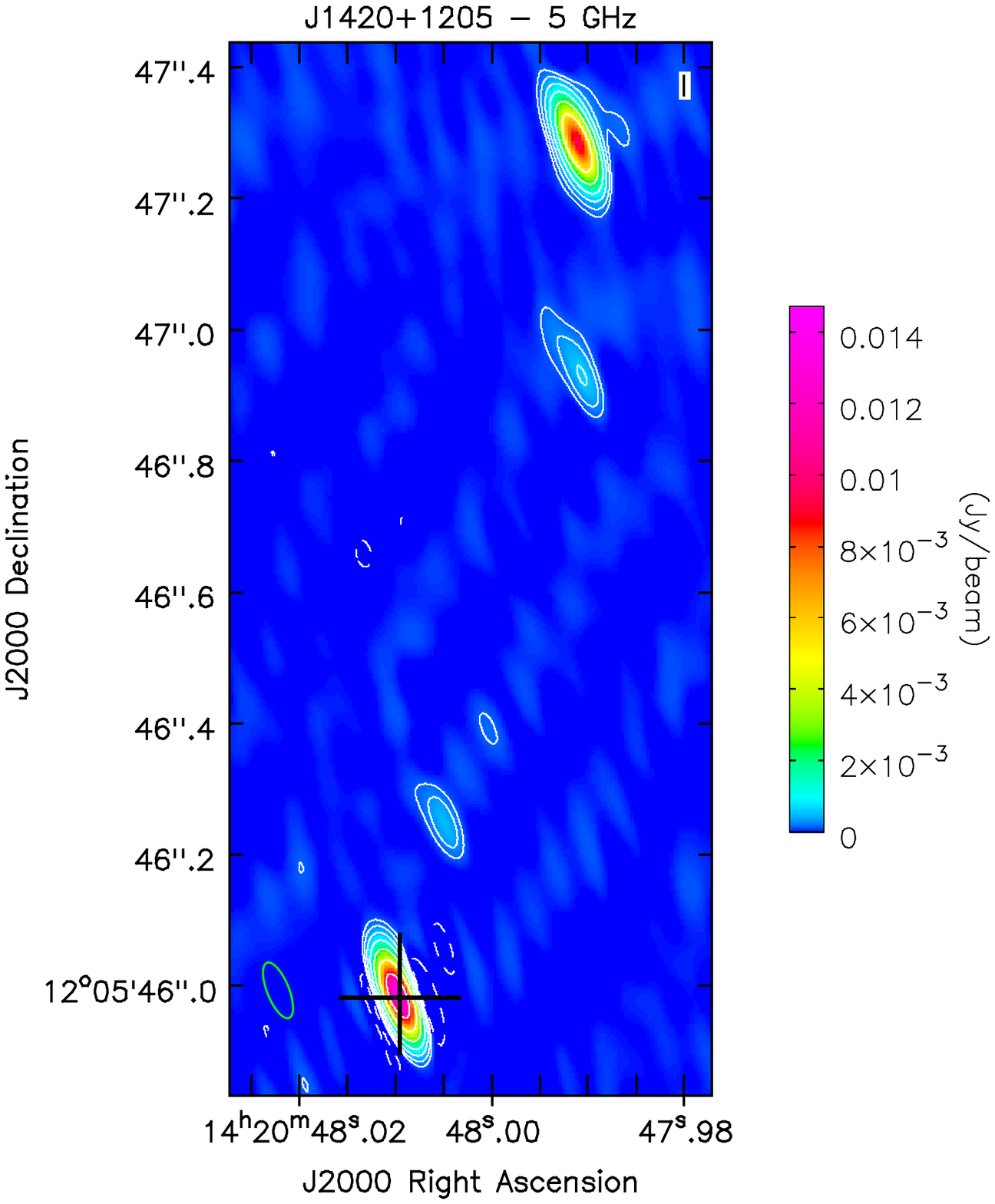}
\end{minipage}
\caption{e-MERLIN images of J1420$+$1205. Black crosses mark the position of the optical source (Gaia DR2, \cite{gaia_dr2}). Restoring beams are shown at the bottom left corner of the images as green ellipses. {\it Left panel:} $1.5$\,GHz map. Peak brightness is $33.3$\,mJy\,beam$^{-1}$. The lowest positive contour level is drawn at $0.3$\,mJy\,beam$^{-1}$ ($5\sigma$), further contour levels increase by a factor of $2$. The beamsize is $0\farcs41 \times 0\farcs14$ at a major axis position angle of $28^\circ$. {\it Right panel:} $5$\,GHz map. Peak brightness is $14.8$\,mJy\,beam$^{-1}$. The lowest positive contour level is drawn at $0.15$\,mJy\,beam$^{-1}$ ($6\sigma$), further contour levels increase by a factor of $2$. The beamsize is $0\farcs09 \times 0\farcs03$ at a major axis position angle of $22^\circ$.}
\label{fig:J1420_emerlin}
\hspace*{\fill}
\end{figure}

At both frequencies, two bright components, and weaker features between them can be detected by e-MERLIN (Fig. \ref{fig:J1420_emerlin}). At $1.5$\,GHz, an additional component can be imaged to the south, which however is below the sensitivity limit in the higher-frequency image. The peak is located in the northern hot spot in the $1.5$-GHz map, while at $5$\,GHz it is in the radio feature corresponding to the centre of the optical galaxy \cite{Cao2017}. 
The features to the south and north of this component detected at $1.5$\,GHz are related to the jet and lobe emissions on either side of the galactic nucleus. The non-detection of the southern lobe at $5$\,GHz is probably due to its steep radio spectrum. The sum of the CLEAN components at $1.5$\,GHz ($74.2 \pm 3.0$\,mJy) is close to the flux density reported in the FIRST catalog ($87.3 \pm 6.6$\,mJy, \cite{FIRST_cat}). Thus we recovered almost all of the radio emission at this frequency (barring possible effects of variability of the compact components). The total extent of the radio-emitting source at $1.5$\,GHz is $\sim 2\farcs5$, corresponding to a $\sim17.3$\,kpc projected size at the redshift of the source. According to the common picture of expanding radio galaxies \cite{arm-length}, the advancing lobe is seen farther away from the centre, thus the arm-length ratio of the brighter to the fainter lobe is larger than unity. This indicates that the brighter, northern hot spot is in the advancing lobe. On the other hand, the bright hot spot emission can be explained as resulting from the interaction between the jet and the surrounding medium, and the fact that it can be seen only on the northern side may indicate asymmetric distribution of the interstellar medium on two sides of the galaxy.

\subsection{J1548$+$3335}

\begin{figure}
\begin{minipage}[t]{.48\textwidth}
\centering
\vspace{0pt}
\includegraphics[bb=25 390 490 660, clip=, width=\textwidth]{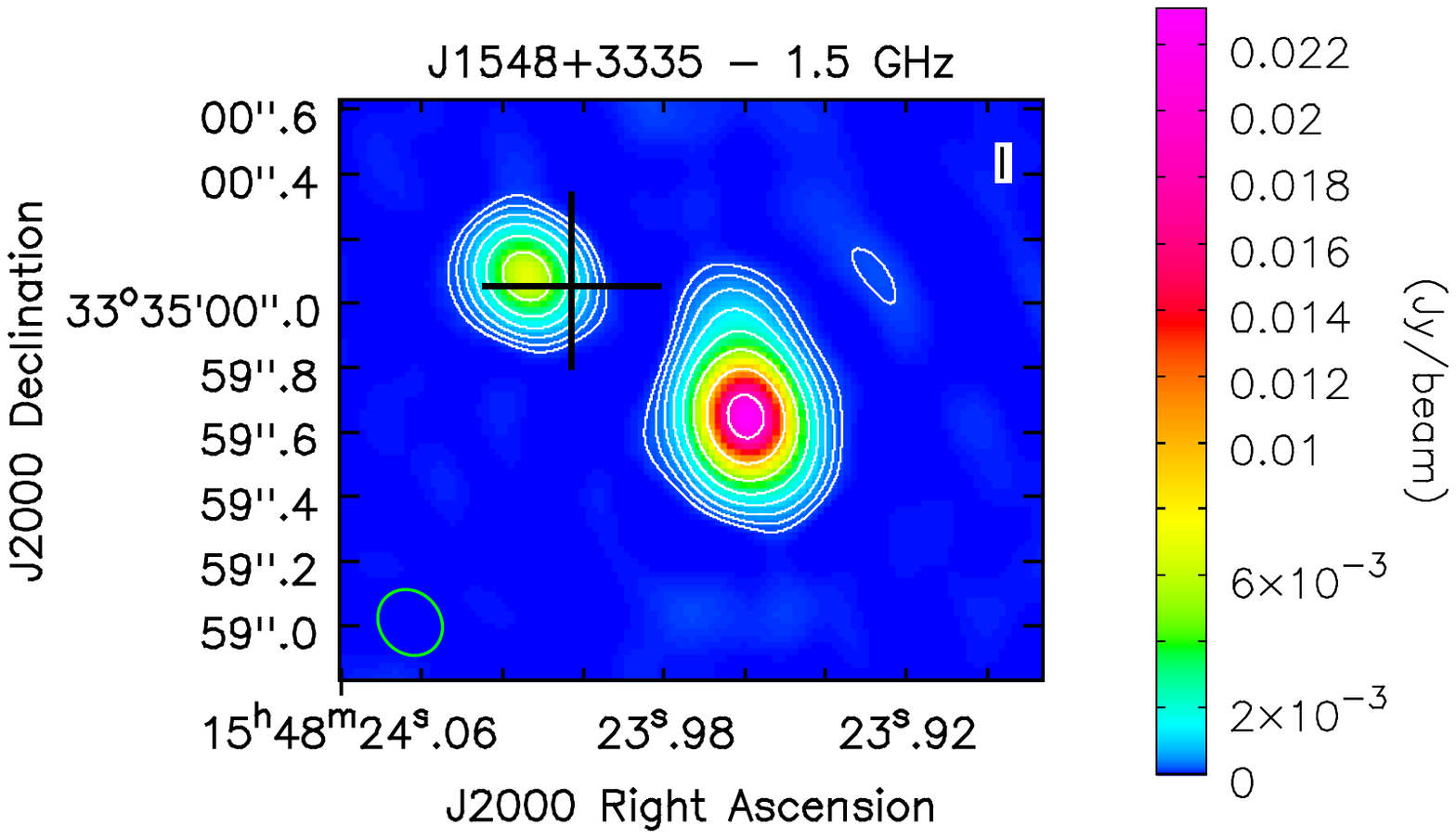}
\end{minipage}
\hfill
\begin{minipage}[t]{.52\textwidth}
\centering
\vspace{0pt}
\includegraphics[bb=20 395 490 660, clip=, width=\textwidth]{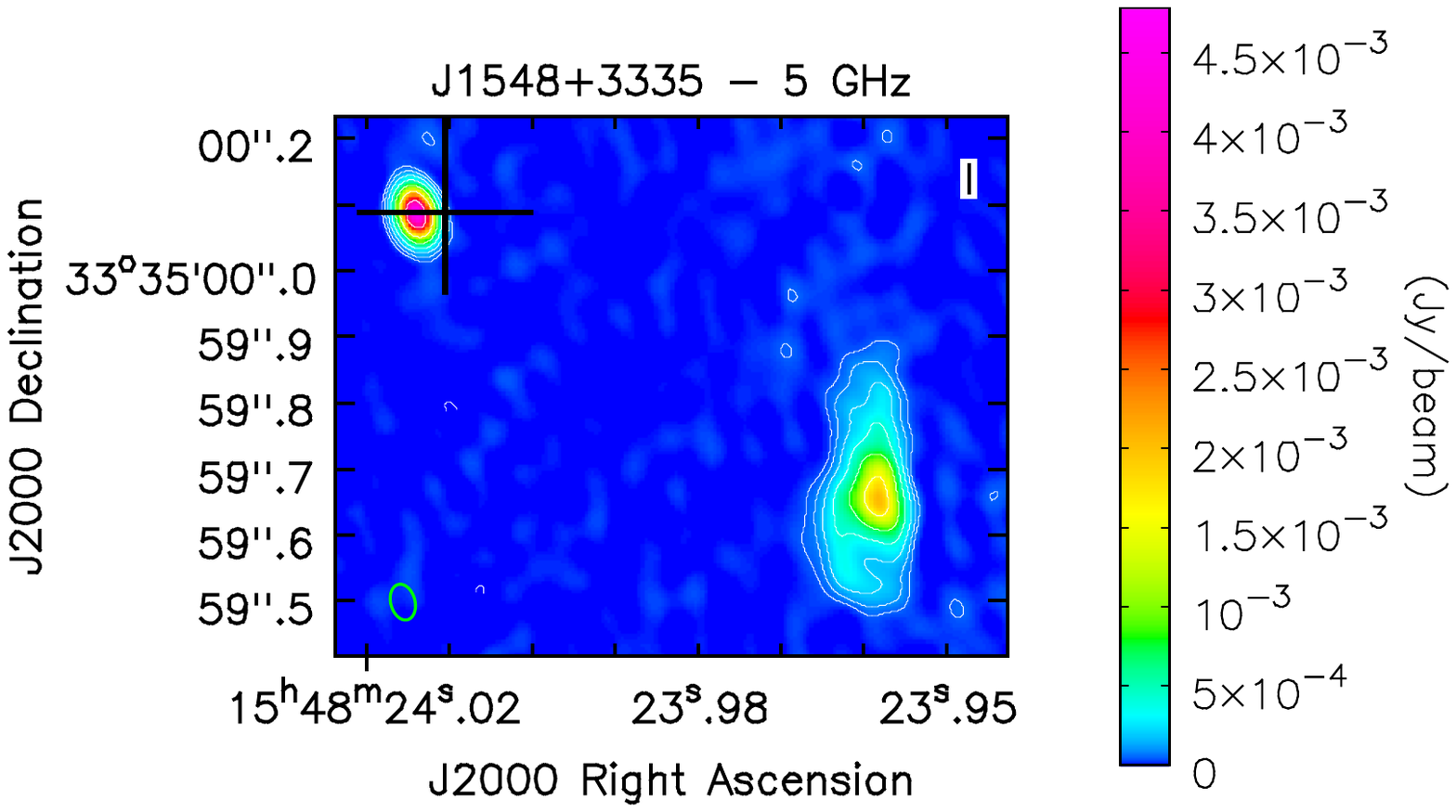}
\end{minipage}
\caption{e-MERLIN images of J1548$+$3335. Black crosses mark the position of the optical source (SDSS DR12, \cite{sdss_dr12}). Restoring beams are shown at the bottom left corner of the images as green ellipses. {\it Left panel:} $1.5$\,GHz map. Peak brightness is $23.2$\,mJy\,beam$^{-1}$. The lowest positive contour level is drawn at $0.2$\,mJy\,beam$^{-1}$ ($5\sigma$), further contour levels increase by a factor of $2$. The beamsize is $0\farcs22 \times 0\farcs19$ at a major axis position angle of $41^\circ$. {\it Right panel:} $5$\,GHz map. Peak brightness is $4.8$\,mJy\,beam$^{-1}$. The lowest positive contour level is drawn at $0.06$\,mJy\,beam$^{-1}$ ($3\sigma$), further contour levels increase by a factor of $2$. The beamsize is $0\farcs06 \times 0\farcs04$ at a major axis position angle of $16^\circ$.}
\label{fig:J1548_emerlin}
\hspace*{\fill}
\end{figure}

Our EVN observations \cite{Coppejans2016} revealed the compact northern feature associated with the centre of the optical galaxy. (The galaxy was not detected by Gaia \cite{gaia_mission}, its SDSS DR12 \cite{sdss_dr12} coordinates agree with the radio coordinates derived from the EVN measurement within the errors.) Our e-MERLIN images (Fig. \ref{fig:J1548_emerlin}) show two features at both frequencies, with the southwest hot spot being brighter at $1.5$\,GHz but the northeast component being more compact and brighter at $5$\,GHz. There is no indication of radio emission to the northwest of the galactic centre down to a $5\sigma$ image noise level of $0.2$\,mJy\,beam$^{-1}$ at $1.5$\,GHz. The flux density recovered at $1.5$\,GHz with e-MERLIN, $39\pm 1$\,mJy is in agreement with the value given in the FIRST catalog, $37.8\pm 1.9$\,mJy \cite{FIRST_cat}. This indicates that there are no extended radio-emitting features at intermediate scales between those probed by FIRST and e-MERLIN. The separation of the two radio features is $\sim 1''$ at $1.5$\,GHz, corresponding to a projected linear size of $6.5$\,kpc at the redshift of the source. The disturbed morphology of the southwest feature may indicate interaction with the dense interstellar material around the jet.

\subsection{J2220$+$0025}

\begin{figure}
\begin{minipage}[t]{.47\textwidth}
\centering
\vspace{0pt}
\includegraphics[bb=40 350 490 720, clip=, width=\textwidth]{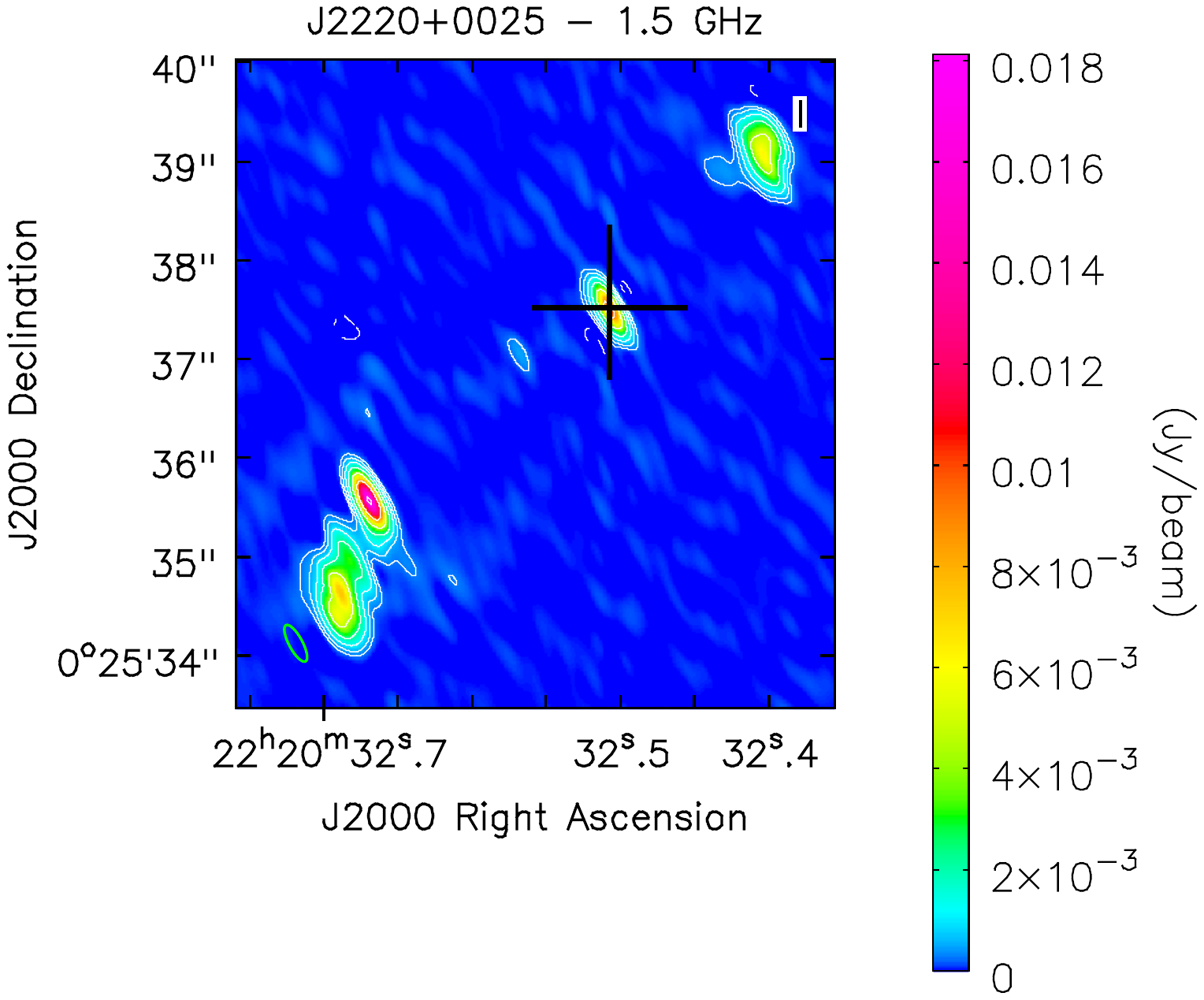}
\end{minipage}
\hfill
\begin{minipage}[t]{.53\textwidth}
\centering
\vspace{0pt}
\includegraphics[bb=40 350 490 660, clip=, width=\textwidth]{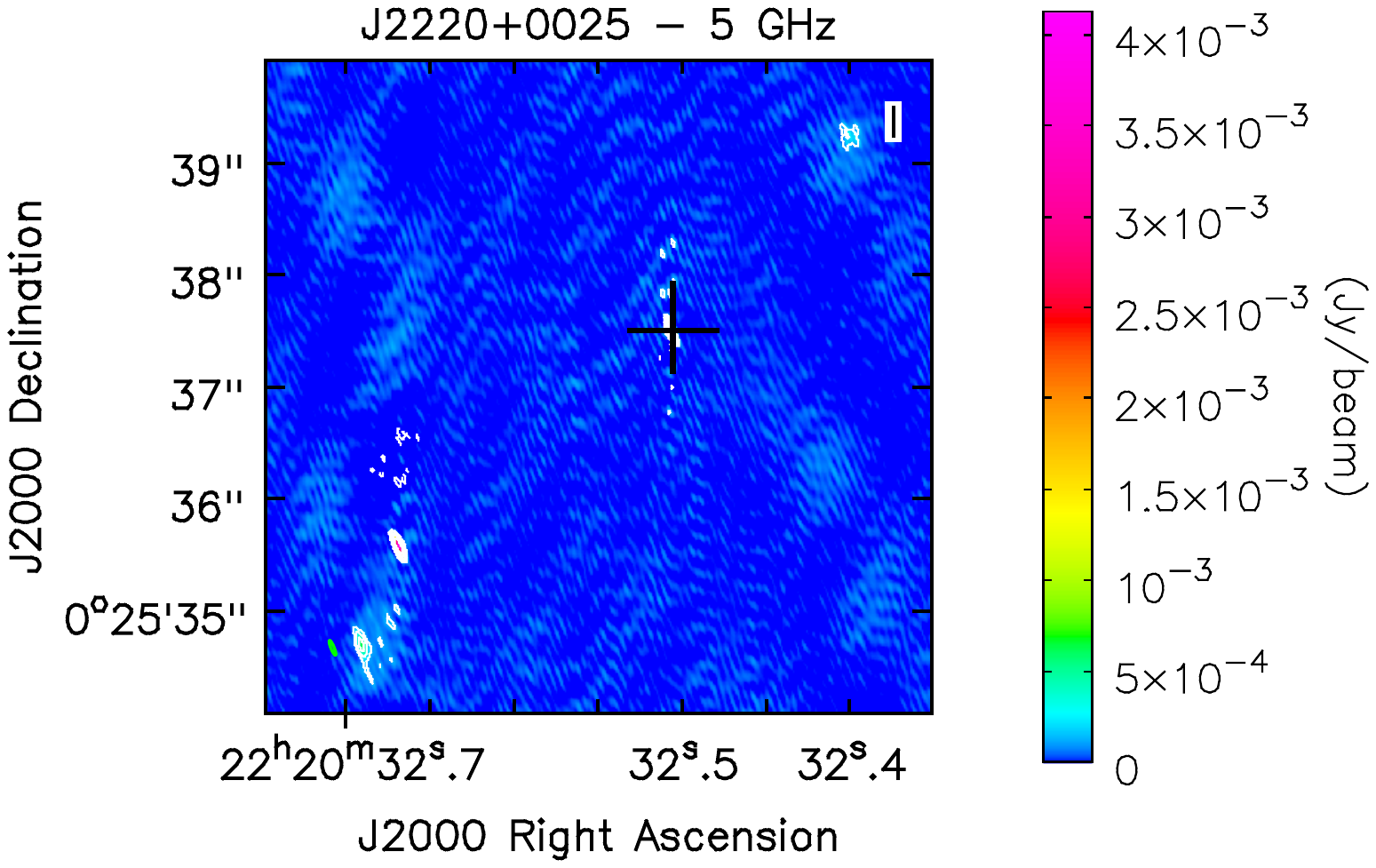}
\end{minipage}
\caption{e-MERLIN images of J2220$+$0025. Black crosses mark the position of the optical source (Gaia DR2, \cite{gaia_dr2}). Restoring beams are shown at the bottom left corner of the images as green ellipses. {\it Left panel:} $1.5$\,GHz map. Peak brightness is $18.2$\,mJy\,beam$^{-1}$. The lowest positive contour level is drawn at $0.3$\,mJy\,beam$^{-1}$ ($5\sigma$), further contour levels increase by a factor of $2$. The beamsize is $0\farcs41 \times 0\farcs14$ at a major axis position angle of $28^\circ$. {\it Right panel:} $5$\,GHz map. Peak brightness is $4.1$\,mJy\,beam$^{-1}$. The lowest positive contour level is drawn at $0.13$\,mJy\,beam$^{-1}$ ($5\sigma$), further contour levels increase by a factor of $2$. The beamsize is $0\farcs14 \times 0\farcs03$ at a major axis position angle of $22^\circ$.}
\label{fig:J2220_emerlin}
\hspace*{\fill}
\end{figure}

In the 1.7-GHz EVN image \cite{Cao2017}, two radio-emitting regions could be modeled, while at $5$\,GHz only the central component could be recovered reliably. The position of the central radio-emitting feature \cite{Cao2017} agrees within the errors with the Gaia DR2 position of the host galaxy \cite{gaia_dr2}. Compared to the data in the FIRST catalog ($92.7\pm6.4$\,mJy, \cite{FIRST_cat}) and those in the VLA-SDSS Stripe 82 survey ($\sim 89.5$\,mJy, \cite{stripe82}), a large fraction of the flux density was resolved out in the EVN observation \cite{Cao2017}. With e-MERLIN, most of this missing flux density could be imaged (the sum of the CLEAN components' flux density is $75.0\pm2.0$\,mJy at $1.5$\,GHz). Our e-MERLIN data revealed a rich morphology (Fig. \ref{fig:J2220_emerlin}), with emission on the jet and counter-jet side at both frequencies. At the lower frequency, the southern hot spot is the brightest feature, while in the $5$-GHz image the peak is located in the central component. 
The total extent of the radio source seen in the 1.5-GHz e-MERLIN image is $\sim 6\farcs5$, corresponding to a projected linear size of $44.2$\,kpc at the redshift of the source. The southern brighter region is located farther away from the central component than the fainter northern one, indicating that the southern arm contains the advancing lobe. On the other hand, similarly to J1420$+$1205, asymmetric distribution of the surrounding material of the galaxy may also cause the asymmetries seen in the brightness distribution. 

\section{Summary}
All three of the targeted $z>4$ radio-loud AGN were detected by e-MERLIN at $1.5$ and $5$\,GHz. The double structures already shown by our previous $1.7$-GHz EVN observations \cite{Cao2017,Coppejans2016} were confirmed and unlike in the EVN observations, we were able to image both features in the three sources even at $5$\,GHz. These originate from the lobes, and the mas-scale compact features detected in them with EVN at $1.7$\,GHz are hot spots. At $5$\,GHz, the peak of the e-MERLIN image coincides with the more compact EVN feature in all sources. These can be associated with the optical centres of the quasar host galaxies. At $1.5$\,GHz, the peak brightness is associated with the hot spots, which is common in Fanaroff-Riley II radio galaxies \cite{fr}. 

The two-sided, radio galaxy-like structure can be further strengthened by the detection of radio emission on the counterjet side in the case of J1420$+$1205 and J2220$+$0025. The total projected linear extent of the radio emission imaged at $1.5$\,GHz is a few kpc in our sources. Their size and luminosity derived from low-resolution FIRST data ($7-20 \times 10^{27}$\,W\,Hz$^{-1}$) are similar to those of medium-sized symmetric objects, which are regarded as younger counterparts of radio galaxies \cite{AnBaan}. However, when comparing to close-by systems, one has to keep in mind that observing frequencies of $1.5$\,GHz and $5$\,GHz correspond to $\sim8$\,GHz and $\sim25$\,GHz in the rest frames of these high-redshift sources. The prominent radio emission from the hot spots at these relatively high frequencies might be related to the more dense interstellar material around these young radio galaxies at early cosmological epochs.

Our results highlight the importance of high-resolution radio interferometric studies in determining the nature of radio-emitting sources. This has enhanced relevance in the case of high-redshift sources, where theoretical predictions are necessarily based on a relatively small sample known at present.

\section*{Acknowledgements}
The EVN is a joint facility of independent European, African, Asian, and North American radio astronomy institutes.  The e-MERLIN is a National Facility operated by the University of Manchester at Jodrell Bank Observatory on behalf of the UK Science and Technology Facilities Council (STFC). This project has received funding from the European Union's Horizon 2020 research and innovation programme under grant agreement No 730562 (RadioNet). K\'EG was supported by the J\'anos Bolyai Research Scholarship of the Hungarian Academy of Sciences. We thank the Hungarian National Research, Development and Innovation Office (OTKA NN110333) for support. HC thanks the Light of West China programme (Grants No. 2016-QNXZ-B-21) and the National Natural Science Foundation of China (Grant No. U1731103). Scientific results presented in this publication are derived from the following EVN and e-MERLIN projects codes: EC052 (PI: D. Cseh), EC054 (PI: H. Cao), and CY5211 (PI: K. \'E. Gab\'anyi). We thank for the usage of the MTA Cloud (cloud.mta.hu) that significantly helped us to achieve the results. This work has made use of data from the European Space Agency (ESA) mission {\it Gaia} (https://www.cosmos.esa.int/gaia), processed by the {\it Gaia} Data Processing and Analysis Consortium (DPAC, https://www.cosmos.esa.int/web/gaia/dpac/consortium). Funding for the DPAC has been provided by national institutions, in particular the institutions participating in the {\it Gaia} Multilateral Agreement.

\bibliographystyle{JHEP}
\bibliography{ref}



\end{document}